\documentclass[aps,pre,reprint,unsortedaddress,showpacs,amsmath,amssymb,twocolumn,preprintnumbers]{revtex4-1}
\usepackage{graphics,graphicx}

\begin{document}
\preprint{Physical Review E}

\title{Large-scale length that determines the mean rate of energy dissipation in turbulence}

\author{Hideaki Mouri}

\author{Akihiro Hori}
\altaffiliation[Also at ]{Meteorological and Environmental Sensing Technology, Inc.}

\author{Yoshihide Kawashima}
\altaffiliation[Also at ]{Meteorological and Environmental Sensing Technology, Inc.}

\author{Kosuke Hashimoto}
\altaffiliation[Also at ]{Meteorological and Environmental Sensing Technology, Inc.}

\affiliation{Meteorological Research Institute, Nagamine, Tsukuba 305-0052, Japan}

\date{August 2, 2012}


\begin{abstract}
The mean rate of energy dissipation $\langle \varepsilon \rangle$ per unit mass of turbulence is often written in the form of $\langle \varepsilon \rangle = C_u \langle u^2 \rangle^{3/2}/L_u$, where the root-mean-square velocity fluctuation $\langle u^2 \rangle^{1/2}$ and the velocity correlation length $L_u$ are parameters of the energy-containing large scales. However, the dimensionless coefficient $C_u$ is known to depend on the flow configuration that is to induce the turbulence. We define the correlation length $L_{u^2}$ of the local energy $u^2$, study $C_{u^2} = \langle \varepsilon \rangle L_{u^2} / \langle u^2 \rangle^{3/2}$ with experimental data of several flows, and find that $C_{u^2}$ does not depend on the flow configuration. Not $L_u$ but $L_{u^2}$ could serve universally as the typical size of the energy-containing eddies, so that $\langle u^2 \rangle^{3/2}/L_{u^2}$ is proportional to the rate at which the kinetic energy is removed from those eddies and is eventually dissipated into heat. The independence from the flow configuration is also found for the two-point correlations and so on when $L_{u^2}$ is used to normalize the scale. 
\end{abstract}

\pacs{47.27.Ak}

\maketitle

\section{Introduction} \label{S1}

Since the kinetic energy of turbulence is transferred from large to small scales and is eventually dissipated into heat, the mean rate of its dissipation per unit mass $\langle \varepsilon \rangle = \nu \sum_{i,j =1}^3 \langle (\partial_{x_i} v_j + \partial_{x_j} v_i)^2 \rangle /2$ is independent of the value of the kinematic viscosity $\nu$. The rate $\langle \varepsilon \rangle$ is instead determined by parameters of the large scales:
\begin{equation}
\label{eq1}
\langle \varepsilon \rangle = C \frac{\langle u^2 \rangle^{3/2}}{L}.
\end{equation}
Here $C$ is a dimensionless coefficient, $\langle u^2 \rangle^{1/2}$ is the root-mean-square fluctuation of the turbulence velocity $v_i$ in some direction $x_i$, and the length $L$ represents the large scales or equivalently the sizes of the energy-containing eddies. The energy of such eddies is of the order of $\langle u^2 \rangle$. Their time scale is of the order of $L/ \langle u^2 \rangle^{1/2}$. As a result, $\langle u^2 \rangle^{3/2}/L$ is of the order of the rate at which their energy is transferred to the smaller eddies. This rate is in turn equal to $\langle \varepsilon \rangle$.

Historically, Eq.~(\ref{eq1}) was found by Taylor in 1935 \cite{t35}. Dryden in 1943 \cite{d43} and Batchelor in 1953 \cite{b53} found that Eq.~(\ref{eq1}) is not inconsistent with experimental data of grid turbulence. Thereafter, Eq.~(\ref{eq1}) has been used as one of the most fundamental laws of turbulence in various studies \cite{f95}, some of which were already described by Landau and Lifshitz in 1959 \cite{ll59}.

The length $L$ is traditionally defined as the correlation length $L_u$ of the velocity $u$ \cite{d43,b53}, which is based on the two-point correlation $\langle u(x+r)u(x) \rangle$:
\begin{subequations}
\label{eq2}
\begin{equation}
\label{eq2a}
L_u = \frac{\int^{\infty}_0 \langle u(x+r)u(x) \rangle d r}{\langle u^2 \rangle}.
\end{equation}
However, with experimental and numerical data of many flows, Sreenivasan \cite{s84,s95,s98} found that $C_u = \langle \varepsilon \rangle L_u / \langle u^2 \rangle^{3/2}$ is not universal but is determined by the flow configuration. That is, $C_u$ depends on the boundary condition, external force, and so on of the turbulence. For the same flow configuration, $C_u$ was almost a constant of order unity when the Reynolds number was high and the energy-containing large scales were separated from the energy-dissipating small scales. These findings have been confirmed experimentally \cite{ap00,pkv02,bla05,mv08} and numerically \cite{kaneda03,goto09,mbsy10}.

The dependence of $C_u$ on the flow configuration implies that $\langle u^2 \rangle^{3/2}/L_u$ is not proportional to the rate at which the energy is removed from the energy-containing eddies. Although $\langle u^2 \rangle$ is not proportional to the total energy $\sum_{i=1}^3 \langle v_i^2 \rangle$ if the turbulence is not isotropic, this is not the sole reason because $C_u$ differs even among isotropic flows \cite{s84,s98,pkv02,bla05,mv08,goto09,kaneda03}. It is concluded that $L_u$ is not proportional to the typical size $L$ of the energy-containing eddies.

This conclusion is important because $L_u$ has been used as a representative of the large scales, not only in studies on Eq.~(\ref{eq1}) but also in many other studies \cite{f95}. It is desirable to find a more universal definition of $L$. Pearson {\it et al.} \cite{pkv02} defined $L$ as a scale that corresponds to the peak of the so-called premultiplied energy spectrum, but their definition has turned out to retain the dependence of $C$ on the flow configuration \cite{bla05}. Here, we define $L$ as the correlation length $L_{u^2}$ of the local energy $u^2$:
\begin{equation}
\label{eq2b}
L_{u^2} = \frac{\int^{\infty}_0 \langle [u^2(x+r) - \langle u^2 \rangle][u^2(x) - \langle u^2 \rangle] \rangle \, d r}
               {\langle (u^2  - \langle u^2 \rangle)^2 \rangle} .
\end{equation}
\end{subequations}
The expectation is that the fluctuations of $u^2$ represented by $L_{u^2}$ could be related to the energy-containing eddies, which are defined to possess the mean energy $\langle u^2 \rangle$. With experimental data of several flows of fully developed turbulence, $C_{u^2} = \langle \varepsilon \rangle L_{u^2} / \langle u^2 \rangle^{3/2}$ is studied over a range of the Reynolds number.

\begin{turnpage}
\begingroup
\squeezetable
\begin{table*}[tbp]
\caption{\label{t1} Experimental conditions and turbulence parameters of grid turbulence G1--G5, boundary layer B1--B6, and jet J1--J6:  incoming flow velocity $U_{\rm wt}$, measurement position $x_{\rm wt}$ and $z_{\rm wt}$, sampling frequency $f$, mean streamwise velocity $U$, kinematic viscosity $\nu$, mean rate of energy dissipation $\langle \varepsilon \rangle = 15 \nu \langle (\partial_x v)^2 \rangle /2$, root-mean-square velocity fluctuations $\langle u^2 \rangle^{1/2}$ and $\langle v^2 \rangle^{1/2}$, correlation lengths $L_{u^n} = \int^{\infty}_0 \langle [u^n(x+r) - \langle u^n \rangle][u^n(x) - \langle u^n \rangle] \rangle \, d r / \langle (u^n  - \langle u^n \rangle)^2 \rangle$ and $L_{v^n} = \int^{\infty}_0 \langle [v^n(x+r) - \langle v^n \rangle][v^n(x) - \langle v^n \rangle] \rangle \, d r / \langle (v^n  - \langle v^n \rangle)^2 \rangle$, and Reynolds numbers Re$_{\lambda_u} = \lambda_u \langle u^2 \rangle^{1/2} /\nu$ and Re$_{\lambda_v} = \lambda_v \langle v^2 \rangle^{1/2} /\nu$ for $\lambda_u = [\langle u^2 \rangle / \langle (\partial_x u)^2 \rangle ]^{1/2}$ and $\lambda_v = [\langle v^2 \rangle / \langle (\partial_x v)^2 \rangle ]^{1/2}$.
}

\begin{ruledtabular}
\begin{tabular}{lccccccccccccccccccc}
\noalign{\smallskip}
                              &
$U_{\rm wt}$                  &
$x_{\rm wt}$                  &
$z_{\rm wt}$                  &
$f$                           &
$U$                           &
$\nu$                         &
$\langle \varepsilon \rangle$ &
$\langle u^2 \rangle^{1/2}$   &
$\langle v^2 \rangle^{1/2}$   &
$L_u$                         &
$L_{u^2}$                     &
$L_{u^3}$                     &
$L_{u^4}$                     &
$L_v$                         &
$L_{v^2}$                     &
$L_{v^3}$                     &
$L_{v^4}$                     &
Re$_{\lambda_u}$              &
Re$_{\lambda_v}$              \\
                 &
m\,s$^{-1}$      &
m                &
m                &
kHz              &
m\,s$^{-1}$      &
cm$^2$\,s$^{-1}$ &
m$^2$\,s$^{-3}$  &
m\,s$^{-1}$      &
m\,s$^{-1}$      &
m                &
m                &
m                &
m                &
m                &
m                &
m                &
m                &
                 &
                 \\
\hline
G1&$ 4$&$ +1.5$&$1.00$&$ 10$&$ 4.31$&$0.141$&$ 0.141 $&$0.236$&$0.231$&$0.164$&$0.0480$&$0.107$&$0.0395$&$0.0413$&$0.0244$&$0.0304$&$0.0195$&$ 153$&$ 104$\\ 
G2&$ 8$&$ +1.5$&$1.00$&$ 24$&$ 8.57$&$0.142$&$ 0.975 $&$0.475$&$0.463$&$0.170$&$0.0483$&$0.110$&$0.0395$&$0.0407$&$0.0249$&$0.0294$&$0.0193$&$ 234$&$ 158$\\
G3&$12$&$ +1.5$&$1.00$&$ 40$&$12.7 $&$0.143$&$ 2.81  $&$0.696$&$0.683$&$0.175$&$0.0490$&$0.114$&$0.0401$&$0.0446$&$0.0237$&$0.0312$&$0.0188$&$ 296$&$ 202$\\
G4&$16$&$ +2.0$&$1.00$&$ 54$&$16.9 $&$0.143$&$ 4.42  $&$0.862$&$0.837$&$0.182$&$0.0535$&$0.119$&$0.0432$&$0.0455$&$0.0265$&$0.0320$&$0.0205$&$ 362$&$ 241$\\
G5&$20$&$ +2.0$&$1.00$&$ 70$&$21.2 $&$0.142$&$ 7.98  $&$1.10 $&$1.06 $&$0.179$&$0.0560$&$0.117$&$0.0458$&$0.0469$&$0.0265$&$0.0327$&$0.0209$&$ 436$&$ 290$\\
B1&$ 2$&$+12.5$&$0.35$&$  4$&$ 1.55$&$0.140$&$ 0.0367$&$0.290$&$0.239$&$0.481$&$0.217 $&$0.331$&$0.180 $&$0.0672$&$0.0844$&$0.0601$&$0.0644$&$ 455$&$ 218$\\
B2&$ 4$&$+12.5$&$0.35$&$ 10$&$ 3.12$&$0.138$&$ 0.244 $&$0.552$&$0.464$&$0.490$&$0.214 $&$0.365$&$0.187 $&$0.0694$&$0.0808$&$0.0653$&$0.0616$&$ 643$&$ 321$\\
B3&$ 8$&$+12.5$&$0.30$&$ 26$&$ 5.93$&$0.143$&$ 2.05  $&$1.18 $&$0.975$&$0.424$&$0.181 $&$0.309$&$0.152 $&$0.0614$&$0.0711$&$0.0564$&$0.0556$&$ 993$&$ 480$\\
B4&$12$&$+12.5$&$0.30$&$ 44$&$ 9.09$&$0.142$&$ 5.21  $&$1.71 $&$1.42 $&$0.433$&$0.203 $&$0.323$&$0.174 $&$0.0604$&$0.0756$&$0.0565$&$0.0586$&$1325$&$ 641$\\
B5&$16$&$+12.5$&$0.25$&$ 60$&$11.3 $&$0.143$&$12.6   $&$2.37 $&$1.96 $&$0.430$&$0.178 $&$0.300$&$0.142 $&$0.0568$&$0.0718$&$0.0520$&$0.0550$&$1621$&$ 780$\\
B6&$20$&$+12.5$&$0.25$&$ 70$&$13.5 $&$0.142$&$19.1   $&$2.99 $&$2.47 $&$0.410$&$0.170 $&$0.300$&$0.131 $&$0.0605$&$0.0690$&$0.0523$&$0.0535$&$2097$&$1017$\\
J1&$ 4$&$+15.5$&$0.45$&$  6$&$ 2.83$&$0.141$&$ 0.0519$&$0.396$&$0.336$&$1.46 $&$0.290 $&$1.12 $&$0.231 $&$0.112 $&$0.137 $&$0.0980$&$0.103$ &$ 709$&$ 362$\\
J2&$ 8$&$+15.5$&$0.40$&$ 16$&$ 5.59$&$0.139$&$ 0.379 $&$0.743$&$0.661$&$1.30 $&$0.263 $&$0.861$&$0.212 $&$0.103 $&$0.126 $&$0.0920$&$0.0985$&$ 932$&$ 522$\\
J3&$16$&$+15.5$&$0.40$&$ 44$&$11.5 $&$0.139$&$ 2.60  $&$1.56 $&$1.36 $&$1.28 $&$0.272 $&$0.873$&$0.228 $&$0.102 $&$0.125 $&$0.0933$&$0.0950$&$1559$&$ 837$\\
J4&$24$&$+15.5$&$0.40$&$ 70$&$17.4 $&$0.139$&$ 7.52  $&$2.34 $&$2.06 $&$1.24 $&$0.275 $&$0.860$&$0.221 $&$0.102 $&$0.127 $&$0.0920$&$0.0949$&$2074$&$1133$\\
J5&$33$&$+15.5$&$0.40$&$ 90$&$22.9 $&$0.139$&$15.4   $&$3.08 $&$2.71 $&$1.28 $&$0.277 $&$0.924$&$0.221 $&$0.105 $&$0.138 $&$0.0924$&$0.0979$&$2505$&$1375$\\
J6&$41$&$+15.5$&$0.40$&$110$&$27.7 $&$0.141$&$23.0   $&$3.93 $&$3.36 $&$1.21 $&$0.269 $&$0.876$&$0.211 $&$0.104 $&$0.129 $&$0.0909$&$0.0993$&$3315$&$1715$\\
\end{tabular}
\end{ruledtabular}
\end{table*}
\endgroup
\end{turnpage}

\section{Experimental data} \label{s2}

The experimental data used here are those of the longitudinal velocity $u$ and the lateral velocity $v$ in grid turbulence (G1--G5), boundary layer (B1--B6), and jet (J1--J6), among each of which the flow configuration was the same but the Reynolds number spans some range. While B1, B4, B6, J1, and J6 are used here for the first time, the others were used also in our recent works \cite{m09,m11}. Their experimental conditions and turbulence parameters are listed in Table~\ref{t1}.

Especially in the estimates of $\langle \varepsilon \rangle$, uncertainties are not avoidable \cite{s95,bla05}. To minimize any of the resulting bias, the data were obtained and processed in the same manner. The details are described below.

\begin{figure}[bp]
\resizebox{7.4cm}{!}{\includegraphics*[6.0cm,3.8cm][15.5cm,26.2cm]{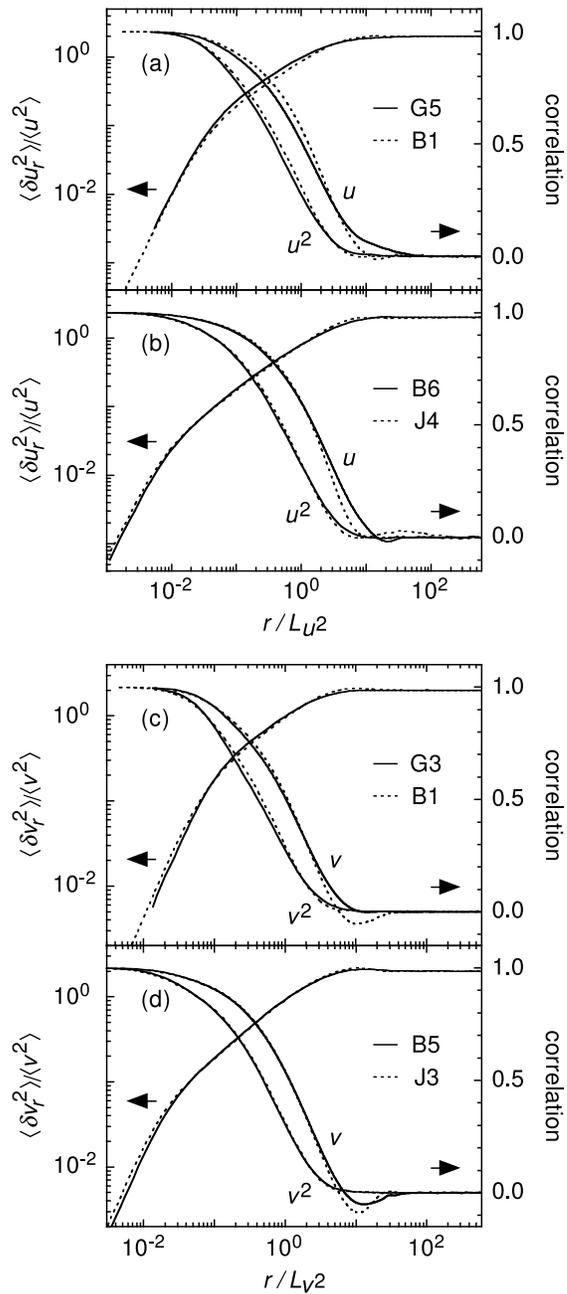}}
\caption{\label{f1} Correlations of $u$ and of $u^2$ and moment $\langle \delta u^2_r \rangle$ as a function of $r/L_{u^2}$ in G5 and B1 and in B6 and J4 (solid and dotted curves). The correlations and the moment are normalized with their values at $r = 0$. Also shown are correlations of $v$ and of $v^2$ and moment $\langle \delta v^2_r \rangle$ as a function of $r/L_{v^2}$ in G3 and B1 and in B5 and J3 (solid and dotted curves).
}
\end{figure} 

\subsection{Experiments} \label{s2a}

The experiments were done in a wind tunnel of the Meteorological Research Institute. We adopt coordinates $x_{\rm wt}$, $y_{\rm wt}$, and $z_{\rm wt}$ in the streamwise, spanwise, and floor-normal directions. The corresponding flow velocities are $U+u$, $v$, and $w$. Here $U$ is the average while $u$, $v$, and $w$ are the fluctuations. The origin $x_{\rm wt} = y_{\rm wt} = z_{\rm wt} = 0$\,m is on the floor center at the upstream end of the test section of the tunnel. Its size was $\delta x_{\rm wt} = 18$\,m, $\delta y_{\rm wt} = 3$\,m, and $\delta z_{\rm wt} = 2$\,m. The cross section $\delta y_{\rm wt} \times \delta z_{\rm wt}$ was the same upstream to $x_{\rm wt} = -4$\,m.

To measure $U+u$ and $v$, we used a hot-wire anemometer. It was composed of a constant temperature system and a crossed-wire probe. The wires were $280$\,$^{\circ}$C in temperature, of platinum-plated tungsten, $5$\,$\mu$m in diameter, $1.25$\,mm in sensing length, $1$\,mm in separation, and oriented at $\pm 45^{\circ}$ to the streamwise direction.  Although the spatial resolution of the probe was not so high, this is not serious as demonstrated in Appendix.

For the grid turbulence G1--G5, a grid was placed at $x_{\rm wt} = -2$\,m across the flow passage to the test section of the wind tunnel. The grid had two layers of rods, with axes in the two layers at right angles. The cross section of the rod was $0.04 \times 0.04$\,m$^2$. The spacing of the rod axes was $0.20$\,m. We set the incoming flow velocity to be $U_{\rm wt} = 4$--$20$\,m\,s$^{-1}$. The measurement was on the tunnel axis, $y_{\rm wt} = 0$\,m and $z_{\rm wt} = 1.00$\,m.

For the boundary layer B1--B6, roughness blocks were placed over the entire floor of the test section. Their size was $\delta x_{\rm wt} = 0.06$\,m, $\delta y_{\rm wt} = 0.21$\,m, and $\delta z_{\rm wt} = 0.11$\,m. The spacing of the block centers was $\delta x_{\rm wt} = \delta y_{\rm wt} = 0.50$\,m. We set the incoming flow velocity to be $U_{\rm wt} = 2$--$20$\,m\,s$^{-1}$. The measurement was in the log-law sublayer at $x_{\rm wt} = +12.5$\,m and $ y_{\rm wt} = 0$\,m, where the boundary layer had the $99$\% velocity thickness of $0.8$\,m.

For the jet J1--J6, we placed a contraction nozzle. Its exit was at $x_{\rm wt} = -2$\,m and was rectangular with the size of $\delta y_{\rm wt} = 2.1$\,m and $\delta z_{\rm wt} = 1.4$\,m. The center was on the tunnel axis. We set the flow velocity at the nozzle exit to be $U_{\rm wt} = 4$--$41$\,m\,s$^{-1}$. The measurement was at $x_{\rm wt} = +15.5$\,m and $y_{\rm wt} = 0$\,m.

The signal of the anemometer was linearized, low-pass filtered, and then digitally sampled. We determined the sampling frequency $f$ as high as possible, on condition that high-frequency noise was not significant in the energy spectrum. The filter cutoff was at $f/2$. We obtained a long record of $1 \times 10^8$ data for $f \le 20$\,kHz or of $4 \times 10^8$ data for $f > 20$\,kHz. Since the variation of air temperature was $\pm 1\,^{\circ}$C at most, the kinematic viscosity $\nu$ was assumed to have been constant.

\subsection{Data processing} \label{s2b}

The temporal fluctuations $u(t_{\rm wt})$ and $v(t_{\rm wt})$ along time $t_{\rm wt}$ were converted to spatial fluctuations of the longitudinal velocity $u(x)$ and of the lateral velocity $v(x)$ along position $x$ through Taylor's hypothesis, $x = -U t_{\rm wt}$. Hereafter, the average $\langle \cdot \rangle$ is taken over the position $x$.

The mean rate of energy dissipation $\langle \varepsilon \rangle$ was calculated as $15 \nu \langle (\partial_x v)^2 \rangle /2$ instead of usual $15 \nu \langle (\partial_x u)^2 \rangle$ by assuming local isotropy, $\langle (\partial_x u)^2 \rangle = \langle (\partial_x v)^2 \rangle /2$. This is because the $v$ measurement was more reliable. The two wires of the crossed-wire probe individually respond to all of the $u$, $v$, and $w$ velocities. Since the measured $u$ velocity corresponds to the sum of the responses of the two wires, it suffers from the $w$ velocity especially at smallest scales. Since the measured $v$ velocity corresponds to the difference of the responses, it does not suffer from the $w$ velocity. We have confirmed the local isotropy by comparing $\delta u_r(x) = u(x+r)-u(x)$ with $\delta v_r(x) = v(x+r)-v(x)$ along the scale $r$. The derivative $\partial _x v$ was estimated as $[ 8v(x+\delta x)-8v(x-\delta x)-v(x+2\delta x)+v(x-2\delta x) ] / 12\delta x$ with the sampling interval $\delta x=U/f$.

We also calculated the Reynolds number Re$_{\lambda}$ for the Taylor microscale $\lambda$ defined as $\lambda_u = [\langle u^2 \rangle / \langle (\partial_x u)^2 \rangle ]^{1/2}$ or as $\lambda_v = [\langle v^2 \rangle / \langle (\partial_x v)^2 \rangle ]^{1/2}$, where the local isotropy was again assumed to estimate $\langle (\partial_x u)^2 \rangle$.

The correlation lengths in Eq.~(\ref{eq2}) were calculated for both the $u$ and $v$ velocities. Usually, the two-point correlation is integrated only up to the scale of its first zero crossing \cite{s95,bla05}. This is to avoid statistical uncertainties at the larger scales, albeit at the expense of any information there. Since our data records are long, we integrated the correlation beyond the scale of the first zero crossing. The convergence of the integration has been confirmed by changing its limit.

Figures~\ref{f1}(a) and \ref{f1}(b) show the two-point correlations $\langle u(x+r)u(x) \rangle$ and $\langle [u^2(x+r) - \langle u^2 \rangle][u^2(x) - \langle u^2 \rangle] \rangle$ for pairs of flows where the configuration is different but the value of Re$_{\lambda_u}$ is similar. The scale $r$ is normalized with the correlation length $L_{u^2}$. While the two curves in each pair are not always identical at $r \gtrsim L_{u^2}$, they are always identical at $r \lesssim L_{u^2}$. Hence, we expect that $L_{u^2}$ does serve as $L$, i.e., typical size of eddies that lie at the top of the energy cascade. This is not expected for $L_u$ because $L_u / L_{u^2}$ is not a constant (see Table \ref{t1}). The $u^2$ correlation decays faster than the $u$ correlation. As a result, $L_{u^2}$ is smaller than $L_u$. This is also the case between the $v$ and $v^2$ correlations in Figs.~\ref{f1}(c) and \ref{f1}(d), although $L_{v^2}$ is larger than $L_v$ in the boundary layer and jet where the $v$ correlation is negative at large $r$. For the same flow configuration, each of $L_u$, $L_{u^2}$, $L_v$, and $L_{v^2}$ has almost the same value in Table \ref{t1}.

\begin{figure}[bp]
\resizebox{7.4cm}{!}{\includegraphics*[5.5cm,3.8cm][15.cm,26.2cm]{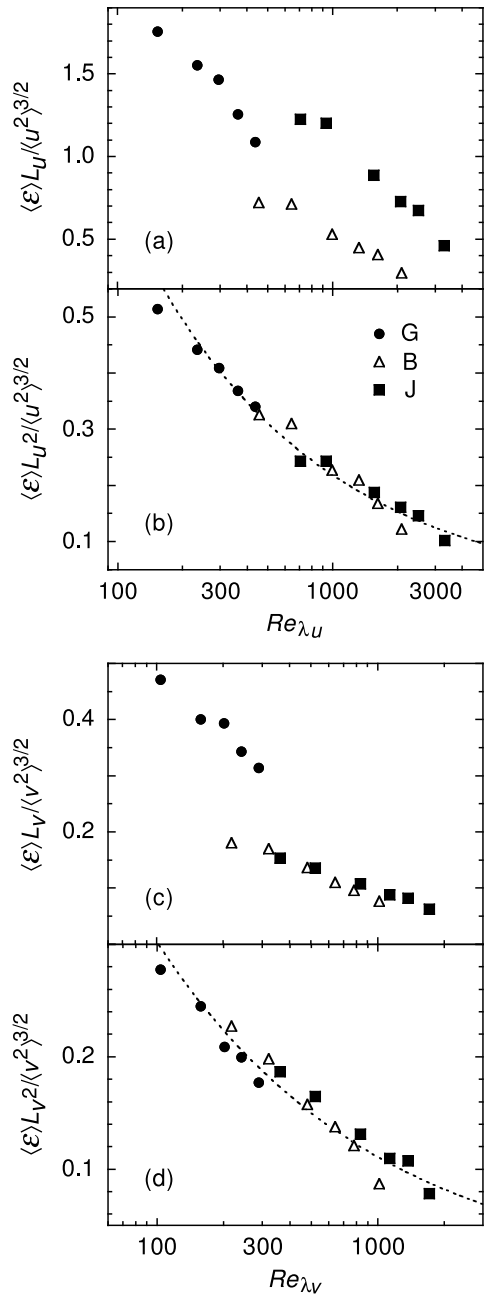}}
\caption{\label{f2} Dimensionless coefficients $C_u = \langle \varepsilon \rangle L_u / \langle u^2 \rangle^{3/2}$ and $C_{u^2} = \langle \varepsilon \rangle L_{u^2} / \langle u^2 \rangle^{3/2}$ as a function of Re$_{\lambda_u}$ and also $C_v = \langle \varepsilon \rangle L_v / \langle v^2 \rangle^{3/2}$ and $C_{v^2} = \langle \varepsilon \rangle L_{v^2} / \langle v^2 \rangle^{3/2}$ as a function of Re$_{\lambda_v}$ in grid turbulence G1--G5 (circles), boundary layer B1--B6 (triangles), and jet J1--J6 (squares). The dotted curve is a fit of $C_{u^2} \propto \mbox{Re}_{\lambda_u}^{-0.51}$ or of $C_{v^2} \propto \mbox{Re}_{\lambda_v}^{-0.43}$.}
\end{figure} 

\section{Results} \label{s3}

Figure~\ref{f2} shows the coefficients $C_u = \langle \varepsilon \rangle L_u / \langle u^2 \rangle^{3/2}$ and $C_{u^2} = \langle \varepsilon \rangle L_{u^2} / \langle u^2 \rangle^{3/2}$ as a function of Re$_{\lambda_u}$ and the coefficients $C_v = \langle \varepsilon \rangle L_v / \langle v^2 \rangle^{3/2}$ and $C_{v^2} = \langle \varepsilon \rangle L_{v^2} / \langle v^2 \rangle^{3/2}$ as a function of Re$_{\lambda_v}$. While the sequences of $C_u$ and $C_v$ do not align among the grid turbulence, boundary layer, and jet in Figs. \ref{f2}(a) and \ref{f2}(c), the sequences of $C_{u^2}$ and $C_{v^2}$ do align in Figs.~\ref{f2}(b) and \ref{f2}(d). Thus, $C_{u^2}$ and $C_{v^2}$ are at least approximately independent of the flow configuration. We favor $L_{u^2}$ and $L_{v^2}$ as the typical size $L$ of the energy-containing eddies.

The sequences of $C_{u^2}$ in Fig.~\ref{f2}(b) appear to align better than those of $C_{v^2}$ in Fig.~\ref{f2}(d). It might follow that $L_{u^2}$ is preferable to $L_{v^2}$, but any conclusion awaits studies of turbulence that is much more anisotropic.

We also consider the correlation lengths of $u^3$, $u^4$, $v^3$, and $v^4$. They are defined similarly to Eq.~(\ref{eq2b}) and are listed in Table~\ref{t1}. As an approximation, we see $L_{u^3} \propto L_u$, $L_{u^4} \propto L_{u^2}$, $L_{v^3} \propto L_v$, and $L_{v^4} \propto L_{v^2}$.  The two lengths in each pair offer almost the same information. Albeit not shown here, the dependence of $\langle \varepsilon \rangle L_{u^3} / \langle u^2 \rangle^{3/2}$ on Re$_{\lambda_u}$ is similar to that of $C_u = \langle \varepsilon \rangle L_u / \langle u^2 \rangle^{3/2}$ in Fig.~\ref{f2}(a), and so on for the other pairs.

Since we have adopted a log scale for Re$_{\lambda}$ in Fig.~\ref{f2}, it is emphasized that $C_u$, $C_{u^2}$, $C_v$, and $C_{v^2}$ are not constants but continue to decrease with an increase in the Reynolds number Re$_{\lambda}$. The same trend is observed in results of the past works \cite{ap00,bla05,mv08,kaneda03,goto09}. Such values of Re$_{\lambda}$ are not yet high enough for complete separation of the large scales from the small scales \cite{ab06}, so that even the energy-containing eddies undergo some energy dissipation in addition to the energy transfer to the smaller eddies, $\propto \langle u^2 \rangle^{3/2} /L$ \cite{mbsy10}. For convenience to Sec.~\ref{s5}, we approximate our results as $C_{u^2} \propto \mbox{Re}_{\lambda_u}^{-\alpha}$ and $C_{v^2} \propto \mbox{Re}_{\lambda_v}^{-\alpha}$ with $\alpha \simeq 1/2$ (dotted curves). The exponent $\alpha$ is nevertheless not a constant and should become zero in the limit of high Re$_{\lambda}$ \cite{s84,s95,s98,pkv02,ap00,bla05,mv08,goto09,kaneda03,mbsy10,l94}.

\begin{figure}[bp]
\resizebox{7.4cm}{!}{\includegraphics*[5.5cm,8.6cm][15cm,21.2cm]{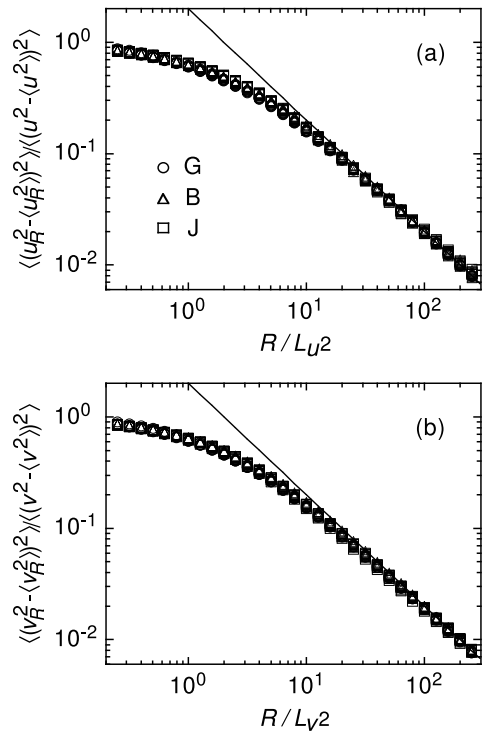}}
\caption{\label{f3} Variances $\langle (u_R^2 - \langle u_R^2 \rangle )^2 \rangle / \langle (u^2 - \langle u^2 \rangle )^2 \rangle$ as a function of $R/L_{u^2}$ and $\langle (v_R^2 - \langle v_R^2 \rangle )^2 \rangle / \langle (v^2 - \langle v^2 \rangle )^2 \rangle$ as a function of $R/L_{v^2}$ in grid turbulence G1--G5 (circles), boundary layer B1--B6 (triangles), and jet J1--J6 (squares). The solid line indicates the relation of Eq.~(\ref{eq4b}).}
\end{figure} 

\section{Discussion} \label{s4}

Having found a possibility that $L_{u^2}$ could serve as the typical size $L$ of the energy-containing eddies, its reasoning is discussed here. The data record of $u(x)$ is divided into segments with length $R$. For each of the segments, the center of which is tentatively defined as $x_{\ast}$, the energy $u^2$ is coarse-grained as
\begin{equation}
\label{eq3}
u_R^2(x_{\ast})=\frac{1}{R} \int^{+R/2}_{-R/2} u^2(x_{\ast}+x) \, d x .
\end{equation}
If $u(x)$ is homogeneous, the mean square of $u_R^2$ around its average $\langle u_R^2 \rangle = \langle u^2 \rangle $ is
\begin{subequations}
\label{eq4}
\begin{align}
\label{eq4a}
&
\langle (u_R^2 - \langle u_R^2 \rangle )^2 \rangle \\
&
= 
\frac{2}{R^2} \int^R_0 (R-r) \langle [u^2(x+r) - \langle u^2 \rangle][u^2(x) - \langle u^2 \rangle] \rangle d r , \nonumber
\end{align}
where $\langle \cdot \rangle$ is used for both the averages over the positions and over the segments \cite{r54}. Also if the correlation of $u^2$ is negligible at $r \gg L_{u^2}$, Eqs.~(\ref{eq2b}) and (\ref{eq4a}) yield a relation of $\langle (u_R^2 - \langle u_R^2 \rangle )^2 \rangle$ to the correlation length $L_{u^2}$ as
\begin{equation}
\label{eq4b}
\langle (u_R^2 - \langle u_R^2 \rangle )^2 \rangle
=
\frac{2L_{u^2}}{R} \langle (u^2 - \langle u^2 \rangle )^2 \rangle
\ \ \
\mbox{at}
\ \ \
R \gg L_{u^2}  .
\end{equation}
\end{subequations}
We could relate $\langle (u_R^2 - \langle u_R^2 \rangle )^2 \rangle$ to the typical size $L$ of the energy-containing eddies, which have been defined to possess the mean energy $\langle u^2 \rangle$. The segment with length $R$ is divided into subsegments with length $N L \ll R$:
\begin{subequations}
\label{eq5}
\begin{equation}
\label{eq5a}
u_R^2= \frac{N L}{R} \sum^{R/N L}_{n = 1} u_{N L}^2(x_n),
\end{equation}
where $x_n$ is the center of the $n$th subsegment \cite{m06}. If $N = 1$, the subsegments are just the energy-containing eddies with the mean energy $\langle u_L^2 \rangle = \langle u^2 \rangle$. The adjacent eddies are correlated with one another, but such a correlation is negligible if we collect a sufficient number of them, $N \gg 1$. In this case, since the variance of $\sum^{R/N L}_{n=1} u_{N L}^2(x_n)$ is $(R/ N L) \langle (u_{N L}^2 - \langle u_{N L}^2 \rangle )^2 \rangle$, we obtain from Eq.~(\ref{eq5a}) as
\begin{equation}
\label{eq5b}
\langle (u_R^2 - \langle u_R^2 \rangle )^2 \rangle
=
\frac{N L}{R} \langle (u_{N L}^2 - \langle u_{N L}^2 \rangle )^2 \rangle .
\end{equation}
Then, Eqs.~(\ref{eq4b}) and (\ref{eq5b}) yield
\begin{equation}
\label{eq5c}
L = \gamma L_{u^2}
\ \ \
\mbox{for}
\ \ \
\gamma = \frac{2 \langle (u^2 - \langle u^2 \rangle )^2 \rangle}{N \langle (u_{N L}^2 - \langle u_{N L}^2 \rangle )^2 \rangle} .
\end{equation}
\end{subequations}
If some fixed values of $\gamma$ and $N$ are used to determine $L$, it is proportional to $L_{u^2}$. The values of $\gamma$ and $N$ are not determined without a further assumption, but we discuss them a little more. Figure~\ref{f3}(a) shows $\langle (u_R^2 - \langle u_R^2 \rangle )^2 \rangle$ in all of our experiments. They are in agreement with Eq.~(\ref{eq4b}) at $R \gtrsim 10^2 L_{u^2}$ (solid line). Since Eq.~(\ref{eq4b}) corresponds to $\gamma =1$ in Eq.~(\ref{eq5c}) with $R = N L$, $N \simeq 10^2$ is enough for $\gamma \simeq 10^0$ to have $L = \gamma L_{u^2}$ independently of the flow. The result also confirms our estimates of $L_{u^2}$. As observed in Fig.~\ref{f3}(b), the same discussion applies to $L_{v^2}$.

This discussion does not apply to $L_u$ or $L_v$. It is true that $L_u$ is related to the mean square of the coarse-grained velocity $U_R(x_{\ast}) = \int^{+R/2}_{-R/2} u(x_{\ast}+x) dx /R$ in a manner similar to Eq.~(\ref{eq4b}):
\begin{equation}
\label{eq6}
\langle U_R^2 \rangle
=
\frac{2L_u}{R} \langle u^2 \rangle 
\ \ \
\mbox{at}
\ \ \
R \gg L_u .
\end{equation}
However, $\langle U_R^2 \rangle$ is not related to the typical size $L$ of the energy-containing eddies so far as they are defined to possess the mean energy $\langle u^2 \rangle$. The reason is $\langle u^2 \rangle > \langle U_R^2 \rangle$ at any $R$. We know of no theory that favors $L_u$ or $L_v$. They rather suffer from too large scales (see Fig.~\ref{f1}). For isotropic turbulence, Batchelor \cite{b53} pointed out that $L_u \propto \int_0^{\infty} k^{-1} E(k) d k$ does not represent the part of the three-dimensional energy spectrum $E(k)$ that makes the major contribution to the energy $\langle u^2 \rangle \propto \int_0^{\infty} E(k) d k$.

Finally, we discuss a relation to the other correlation lengths. There have been obtained $L_{u^4} \propto L_{u^2}$ and $L_{v^4} \propto L_{v^2}$ (see Sec.~\ref{s3}). It might follow that each of $L_{u^2}$ and $L_{v^2}$ as $L$ could be replaced with an arbitrary correlation length based on the absolute value of the velocity, $L_{\vert u \vert^n}$ or $L_{\vert v \vert^n}$. Even in this case, we prefer $L_{u^2}$ and $L_{v^2}$. The energies $u^2$ and $v^2$ are the most fundamental quantities among those that could be related to the absolute value $\vert u \vert$ or $\vert v \vert$ of the velocity.

\begin{figure}[tbp]
\resizebox{7.4cm}{!}{\includegraphics*[5.5cm,2.7cm][15.0cm,28.2cm]{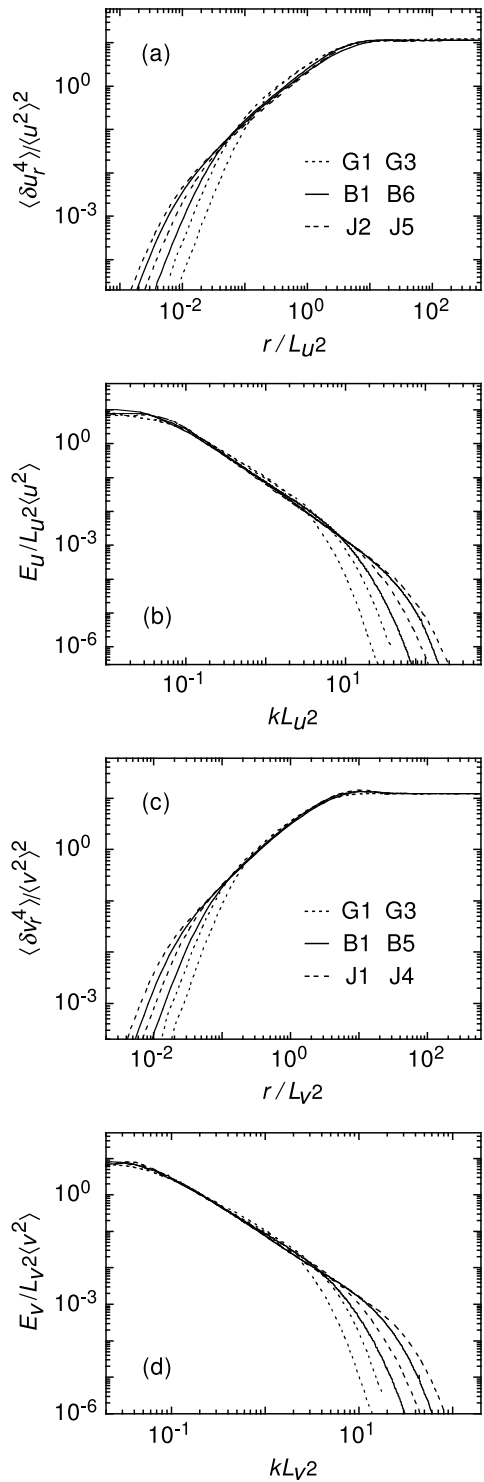}}
\caption{\label{f4} Moment $\langle \delta u_r^4 \rangle / \langle u^2 \rangle^2$ as a function of $r/L_{u^2}$ and spectrum $E_u/L_{u^2} \langle u^2 \rangle$ as a function of $k L_{u^2}$ in G1 and G3 (dotted curves), B1 and B6 (solid curves), and J2 and J5 (dashed curves). The unit of $k$ is m$^{-1}$ instead of usual rad\,m$^{-1}$. Also shown are moment $\langle \delta v_r^4 \rangle / \langle v^2 \rangle^2$ as a function of $r/L_{v^2}$ and spectrum $E_v/L_{v^2} \langle v^2 \rangle$ as a function of $k L_{v^2}$ in G1 and G3 (dotted curves), B1 and B5 (solid curves), and J1 and J4 (dashed curves).}
\end{figure} 
\begin{figure}[tbp]
\resizebox{7.4cm}{!}{\includegraphics*[5.5cm,15.1cm][15.0cm,26.2cm]{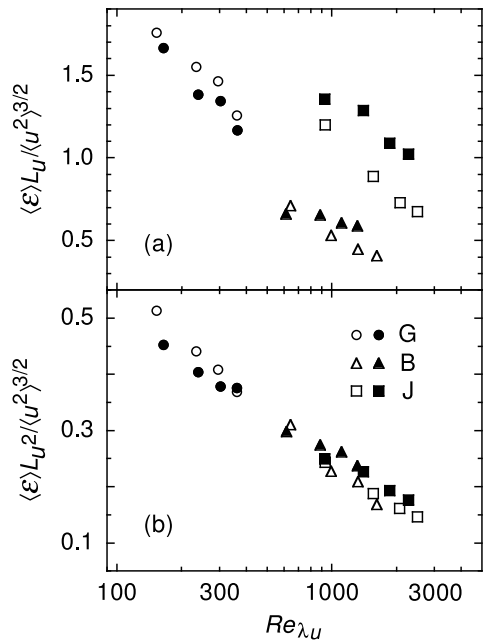}}
\caption{\label{f5} Dimensionless coefficients $C_u = \langle \varepsilon \rangle L_u / \langle u^2 \rangle^{3/2}$ and $C_{u^2} = \langle \varepsilon \rangle L_{u^2} / \langle u^2 \rangle^{3/2}$ as a function of Re$_{\lambda_u}$ in grid turbulence G1--G4 (circles), boundary layer B2--B5 (triangles), and jet J2--J5 (squares). The filled and the open symbols denote data obtained with the single- and the crossed-wire probes.}
\end{figure} 

\section{Concluding Remarks}  \label{s5}

For the mean rate of energy dissipation written in the form of $\langle \varepsilon \rangle = C \langle u^2 \rangle^{3/2}/L$, it is traditional to define $L$ as the correlation length $L_u$ of the velocity $u$. However, $C_u = \langle \varepsilon \rangle L_u / \langle u^2 \rangle^{3/2}$ depends on the flow configuration that is to induce the turbulence \cite{s84,s95,s98,ap00,pkv02,bla05,mv08,goto09,kaneda03}. We have defined $L$ as the correlation length $L_{u^2}$ of the local energy $u^2$, studied $C_{u^2} = \langle \varepsilon \rangle L_{u^2} / \langle u^2 \rangle^{3/2}$ for several flows, and found that $C_{u^2}$ does not depend on the flow configuration. Not $L_u$ but $L_{u^2}$ could serve universally as the typical size $L$ of the energy-containing eddies, so that $\langle u^2 \rangle^{3/2}/L_{u^2}$ is proportional to the rate at which their kinetic energy is transferred to the smaller eddies and is eventually dissipated into heat. The energy-containing eddies, if defined to possess the mean energy $\langle u^2 \rangle$, correspond to spatial fluctuations of $u^2$ represented by $L_{u^2}$ rather than to those of $u$ represented by $L_u$. This is also the case for the correlation length $L_{v^2}$.

The present approach does not yield the value of the constant $\gamma$ for $L = \gamma L_{u^2}$ or $L = \gamma L_{v^2}$. To determine this value, some assumption should be necessary.

We find the independence from the flow configuration also for other statistics when the scale is normalized with $L_{u^2}$ or $L_{v^2}$.  Although smallest scales suffer significantly from the energy dissipation that depends on the Reynolds number Re$_{\lambda}$, the examples include two-point correlations and moments of the velocity difference $\langle \delta u_r^n \rangle / \langle u^2 \rangle^{n/2}$ at the scale of $r \lesssim L_{u^2}$ in Figs.~\ref{f1} and \ref{f4}. Thus, $L_{u^2}$ is proportional to the typical size of eddies that lie at the top of the energy cascade. The equivalent independence is found for the longitudinal energy spectrum $E_u$ at wave number $k$ by normalizing them as $E_u /L_{u^2} \langle u^2 \rangle$ and $k L_{u^2}$ in Fig.~\ref{f4}. Also independent is the variance of the coarse-grained energy $u_R^2$ in Fig.~\ref{f3}. In fact, our recent work \cite{m11} has reproduced its whole distribution at the scale of $R \gtrsim L_{u^2}$ by using $L = 4 L_{u^2}$, i.e., $\gamma = 4$, for Eq.~(\ref{eq5c}). Therefore, $L_{u^2}$ and $L_{v^2}$ are fundamental units of the energy-containing large scales.

The existing discussions on $C = \langle \varepsilon \rangle L / \langle u^2 \rangle^{3/2}$ often assume that $C$ is independent of the Reynolds number Re$_{\lambda}$ \cite{b53,f95,ll59}. They have to be corrected at Re$_{\lambda} \lesssim 10^3$, where we see $C \propto \mbox{Re}_{\lambda}^{-\alpha}$ in Fig.~\ref{f2}. The large-scale Reynolds number $L \langle u^2 \rangle^{1/2} / \nu$ is corrected as $\propto \mbox{Re}_{\lambda}^{2-\alpha}$. The number of degrees of freedom $(L/\eta)^3$, where $\eta = ( \nu^3 / \langle \varepsilon \rangle )^{1/4}$ is the Kolmogorov length, is corrected as $\propto \mbox{Re}_{\lambda}^{9/2-3\alpha}$. Also if, say, Loitsyansky's invariant holds as $\propto L^5 \langle u^2 \rangle$ in decaying isotropic turbulence, $\partial_t \langle u^2 \rangle \propto - \langle \varepsilon \rangle$ yields the decay law $\langle u^2 \rangle \propto t^{-(10-5\alpha)/(7-5\alpha)}$. These relations could have universal coefficients for $L \propto L_{u^2}$ or $L \propto L_{v^2}$.

We also remark on the $K$-$\varepsilon$ model \cite{ls74}, which calculates the eddy viscosity at each position as $\nu_T = c K^2 / \langle \varepsilon \rangle$. Here $K^2 = \sum_{i=1}^3 \langle v_i^2 \rangle /2$ and $\langle \varepsilon \rangle$ are ensemble averages. Defined as $c = \tilde{c} \tilde{C}$ for $\tilde{c} = \nu_T / K^{1/2} L$ and $\tilde{C} = \langle \varepsilon \rangle L / K^{3/2}$, the constant $c = 0.09$ applies to various flows. Then, if $L$ could be defined as the correlation length of some local energy so that $\tilde{C}$ is independent of the flow configuration, a constant value of $\tilde{c}$ applies to the various flows.

Since our data set is not large, we confine ourselves to pointing out the possibility that $L_{u^2}$ or $L_{v^2}$ could serve as $L$. This has to be confirmed in future with a large set of experimental or numerical data. Nevertheless, for this promising possibility, it is already certain that $L_{u^2}$ and $L_{v^2}$ are preferable to the traditional length $L_u$ as the typical size $L$ of the energy-containing eddies or equivalently as the representative of the large scales.

\begin{acknowledgments}
This work was supported in part by KAKENHI Grant No. 22540402. We thank M. Takaoka and T. Matsumoto for stimulating discussions.
\end{acknowledgments}

\appendix*
\section{SUPPLEMENTARY EXPERIMENTS} \label{app}

The spatial resolution of our crossed-wire probe was not so high. We supplementarily used a single-wire probe to measure the streamwise velocity $u$ for the configurations of G1--G4, B2--B5, and J2--J5. The wire parameters were the same as those of the crossed-wire probe. We estimated $\langle \varepsilon \rangle$ as $15 \nu \langle (\partial_x u)^2 \rangle$. It has been ascertained that the values of $-\langle (\partial_x u )^3 \rangle / \langle (\partial_x u )^2 \rangle^{3/2}$ agree with those of experimental and numerical data in the literature, $0.5$--$0.6$ at Re$_{\lambda} \simeq 10^2$--$10^3$ \cite{sa97}. The values smaller by $0.1$--$0.3$ were obtained with our crossed-wire probe.

Figure \ref{f5} compares the results for the single-wire probe (filled symbols) with those for the crossed-wire probe (open symbols). They are in agreement with each other. The exception is $C_u = \langle \varepsilon \rangle L_u / \langle u^2 \rangle^{3/2}$ in the jet where $L_u$ is different by $15$--$25$\,\% (squares). This is because not identical were the $u$ velocities measured with the single- and crossed-wire probes, which were set differently to the traverse system of the wind tunnel. The contamination with the $w$ velocity was also different (see Sec.~\ref{s2b}). Since an alignment of the sequences of $C_{u^2} = \langle \varepsilon \rangle L_{u^2} / \langle u^2 \rangle^{3/2}$ is again observed in Fig.~\ref{f5}(b), the spatial resolution of our crossed-wire probe is not serious to the estimates of $\langle \varepsilon \rangle$, albeit serious to those of $\langle \varepsilon ^n \rangle$ for $n \ge 2$, and hence is not serious to our result that $C_{u^2}$ is independent of the flow configuration at least as a good approximation.

\end{document}